\documentclass[conference]{IEEEtran}
\ifCLASSINFOpdf
\else
\fi

\usepackage[normalem]{ulem}
\useunder{\uline}{\ul}{}
\usepackage{amssymb}
\usepackage{array}
\usepackage{multirow}
\usepackage[table,xcdraw]{xcolor}
\usepackage{graphicx}
\begin{document}
%
\title{An Exploratory Study of Malicious Link Posting on Social Media Applications}


\author{\IEEEauthorblockN{\hspace{-1.93cm}Muhammad Hassan}
\IEEEauthorblockA{\hspace{-1.93cm}University of Illinois at Urbana Champaign\\
\hspace{-1.93cm}mhassa42@illinois.edu}
\and
\IEEEauthorblockN{\hspace{-1.1cm}Mahnoor Jameel}
\IEEEauthorblockA{\hspace{-1.1cm}University of Illinois at Urbana Champaign\\
\hspace{-1.1cm}mjameel2@illinois.edu}

\and
\IEEEauthorblockN{\hspace{2.2cm} Masooda Bashir}
\IEEEauthorblockA{\hspace{2.2cm} University of Illinois at Urbana Champaign\\
\hspace{2.2cm} mnb@illinois.edu}
}




%


\IEEEoverridecommandlockouts
\makeatletter\def\@IEEEpubidpullup{6.5\baselineskip}\makeatother
\IEEEpubid{\parbox{\columnwidth}{
    Symposium on Usable Security and Privacy (USEC) 2023 \\
    27 February 2023, San Diego, CA, USA \\
    ISBN 1-891562-91-6 \\
    https://dx.doi.org/10.14722/usec.2023.234399 \\
    www.ndss-symposium.org, https://www.usablesecurity.net/USEC/
}
\hspace{\columnsep}\makebox[\columnwidth]{}}

\maketitle

\begin{abstract}

Social network platforms are now widely used as a mode of communication globally due to their popularity and their ease of use.  Among the various content-sharing capabilities made available via these applications, link-sharing is a common activity among social media users. While this feature provides a desired functionality for the platform users, link sharing enables attackers to exploit vulnerabilities and compromise users’ devices. Attackers can exploit this content-sharing feature by posting malicious/harmful URLs or deceptive posts and messages which are intended to hide a dangerous link. However, it is not clear how the most common social media applications monitor and/or filter when their users share malicious URLs or links through their platforms. To investigate this security vulnerability, we designed an exploratory study to examine the top five android social media applications’ performance when it comes to malicious link sharing. The aim was to determine if the selected applications had any filtering or defenses against malicious URL sharing. Our results show that most of the selected social media applications did not have an effective defense against the posting and spreading of malicious URLs. While our results are exploratory, we believe our study demonstrates the presence of a vital security vulnerability that malicious attackers or unaware users can use to spread harmful links. In addition, our findings can be used to improve our understanding of link-based attacks as well as the design of security measures that usability into account.

\end{abstract}

\section{\textbf{Introduction}}
\label{sec:introduction}
Social media and networking platforms have made communication and sharing more convenient, hence increasing connectivity in the online world. In 2021, 72\% of the US adult population reported using social media such as Facebook, Twitter, Instagram, etc \cite{pew}. People share their regular life updates and opinions with their friends, family, and colleagues by posting and exchanging messages on social media applications. These platforms also allow their users to share external content among their social networks of friends and followers as well by posting external links. These functionalities make social media usage meaningful and interactive for their users. As a result, social platforms have not only kept users engaged and claim a regular active user base, but popular platforms like Facebook, Twitter, and WhatsApp also have boasted a growth in their user bases.~\cite{clement_2022}~\cite{dixon_2022}~\cite{ceci_2022}.

The Online Social Networks (OSNs) ecosystem is extremely powerful due to its ever-changing nature and fast ease of communication. It adapts to the demands of users and situations. For example, it can be used for knowledge and information dissipation overcoming modern-day censorship ~\cite{rubin2020evolution}, spreading propaganda at the time of conflict ~\cite{nikolayenko2019framing}, and providing assistance in day-to-day life in the face of natural calamity ~\cite{saad2020towards}. Link sharing is a particularly popular feature used by people online. Users share links to external content online, such as posting news links from external sources, sharing photos, videos, stories or disappearing pictures \cite{dixon_2022_share} \cite{giglietto2020coordinated}. Yet, this variability of features also provides a breeding ground for bad actors to take advantage of various usability features and exploit social media for malicious activities.

Prior research demonstrates mobile users tend to ignore posted security warnings, and even more technically skilled users risked privacy and security intrusions despite their awareness of potential risks ~\cite{felt2012android}~\cite{barth2019putting}. This risky behavior could be partly attributed to people’s perception of trust in social networks because they may perceive as strong association with these online communities as their offline ones \cite{ellison2007benefits} \cite{antoci2019civility}. The presence of friends, family and loved ones on social networks can also be attributed to users’ trust in social networks \cite{sherchan2013survey}]. Therefore, this sense of trust can lead to negligent security behaviors among social media users. More importantly, malicious actors, like hackers and scammers, take advantage of such negligence to deceive susceptible people into clicking on malicious links which can lead to monetary loss, and business and reputation damages.

In addition, the popularity of OSNs presents a wide attack surface for online scammers and attackers. Attackers can gain from this opportunity by exploiting the limitations of security design. There are several ways for attackers to motivate users into opening a malicious website. Prior research has explored various attacks stemming from the usability features of social networks. For example, attackers have engineered malicious links to steal user credentials and authentication information from popular social media channels like Facebook, Twitter, etc (e.g. ~\cite{7841038}\cite{phishing}\cite{bbc_news_2020}). Attackers can also gain access to a user’s device by installing malware, without their knowledge \cite{wazid2019mobile}, by spreading deceitful links on social networks. The threat model varies from impersonating a financial entity, or job opportunity, offering discounts and shopping vouchers, to claiming software updates. Attackers using link-based attacks aim to obtain for personally identifiable information \textit{PIIs} such as users' email, social security and demographic information, banking credentials, or corporate and business data. Owing to the interconnected nature of Online Social Networks (OSNs), scammers and hackers ultimately have an ideal attack vector for spreading harmful links which can result in important information loss for users., Therefore, it is critical to investigate if popular social media applications allow users to share malicious links (URLs) via their content-sharing functionality.



In this study, we conducted an empirical investigation to understand how social media applications monitor and allow the sharing of malicious URLs by their users. The goal of this research was to gain a better understanding of the security measures implemented by social network platforms when it comes to malicious URLs, and to investigate if the platforms intercept and check for links that are potentially harmful. We focused our study on popular social media applications on the Android platform.

\vspace{2pt}\noindent To summarize, our paper makes the following contributions

\vspace{3pt}\noindent$\bullet$ We conducted an analysis of the link filtering mechanisms in place on social media platforms and how they impact the user experience when sharing links online.

\vspace{3pt}\noindent$\bullet$ We examined the methods by which malicious actors may bypass these filtering mechanisms in order to post malicious links on social communication channels and media.


\vspace{3pt}\noindent$\bullet$ We provided recommendations and discussed potential future extensions of the work which can provide insights into usable security measures that social media applications can implement to reduce malicious link posting and link-based attacks on social platforms.

    
    


\section{\textbf{Background}}
\label{sec:background}
\vspace{5mm}
\begin{figure*}[!t]
    \centering
    \includegraphics[ width=1.0\textwidth]
    {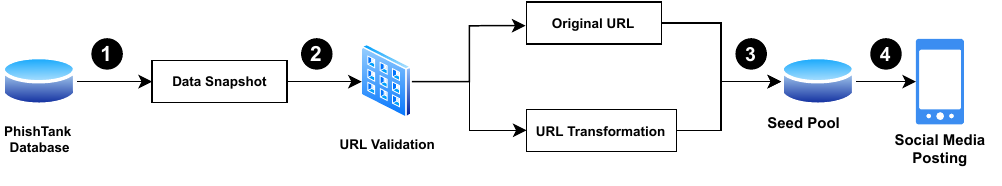}

    \caption{Design and workflow of our pipeline. In \textbf{Process \textit{1} \& \textit{2}} we collect flagged URLs from PhishTank and verify them being harmful from \textit{Google Safe Browsing} and \textit{Virus Total}. Next, we create \textit{transformed} version of every original link and store them in \textbf{Process \textit{3}} and post using the test accounts at \textbf{Process \textit{4}}.}
    
    \label{fig:pipeline}
\end{figure*}

In this section, we discuss the rationale behind our exploratory experiment. We also discuss relevant background on how users are susceptible to the dangers of phishing and specifically link-based attacks as well as what detection mechanisms are often put in place in digital spaces and social networks to deal with this threat.

\subsection{\textbf{Phishing on Social Networks}}

Phishing is a type of cyber attack that involves social engineering techniques to deceive users into clicking on a compromised message or post, usually resulting in the reveal of sensitive and personal information, installation of malware and/or a system being hijacked by the attackers. Online social networks are being used to share personal information online and communicate with friends and family, hence making them more prone to phishing attacks.

Recent research has brought attention to the growing issue of phishing on online social networks. A study by Alharbi and colleagues \cite{alharbi2022security} found that phishing attacks on online platforms are increasing both in frequency and complexity. Their research revealed attackers use a variety of tactics, including creating fake profiles, sharing malicious links, and impersonating reputable brands, to deceive users. Additionally, the study discovered users often had difficulty distinguishing between legitimate and phishing messages, making them vulnerable to these attacks.

A comprehensive survey conducted by K. L. Chiew et al. \cite{chiew2018survey} examined the various phishing vectors and technical approaches employed by attackers. The study found that phishers utilize a wide range of techniques to deceive users, including exploiting browser vulnerabilities, cross-site scripting (XSS) attacks, and clickjacking techniques to send direct messages and post comments containing malicious links on users' profiles. Additionally, the study determined attackers often employ social engineering tactics such as creating a sense of urgency or fear to increase the likelihood of users falling for the attack.

Additionally, a study by F. A. Turjman and R. Salama \cite{al2021cyber} explored the phishing risk on mobile social networks. The mobile versions of social networks are also susceptible to phishing attacks, with attackers utilizing mobile-specific methods such as SMS and MMS messages and mobile applications to execute their attacks.

Link-based attacks involve malicious links embedded in the content of an email, online message, or post. phishing is one of the oldest link-based attacks and still the most effective threat vector on social networks ~\cite{apwg_2022}. Popular examples include receiving unsolicited emails with links redirecting to malicious websites. More recently, attackers have evolved and incorporated social engineering to capitalize on the susceptibility of the end-user on a platform, which shows security design negligence on part of the social platforms~\cite{bakhshi2017social}. Phishing takes advantage of users’ psychological practices, by inducing a sense of urgency and fear to persuade users subconsciously into opening the link ~\cite{sibrian2020sensitive}\cite{sharma2021phishing}. Given the psychological nature of the attack vector, a study by Ion et al. \cite{ion2015no} highlights that even security experts are susceptible to falling victim to phishing attacks, often as a result of overconfidence. The research emphasizes that despite the knowledge and experience of security experts, the psychological manipulation tactics used by attackers can still be effective in tricking them. 

\vspace{1cm}
\subsection{Phishing Detection on Social Networks}

Online social networks are vulnerable to phishing attacks, which use social engineering techniques to trick individuals into revealing sensitive information. Detection of phishing in online social networks is a challenging task, but various techniques such as blacklist URL databases and rule-based heuristic link analysis have been proposed and employed by social networks to mitigate these threats.
Jiwon Hong et. al. \cite{hong2020phishing} present a new approach to detecting phishing URLs based on the use of lexical features and a blacklist of known phishing domains. This approach was able to achieve an F-1 score of 0.84 in detecting phishing URLs. Additionally, the study also emphasized that the use of a blacklist of known phishing domains was particularly effective in detecting phishing URLs.

The use of URL blacklisting services is a well-established technique for detecting phishing links. In their study, S. Bell and P. Komisarczuk \cite{bell2020analysis} conducted an analysis of three popular blacklists for phishing: Google Safe Browsing, OpenPhish, and PhishTank. Among the three services, Google Safe Browsing had the highest coverage of phishing URLs compared to OpenPhish and PhishTank. The study also highlighted the importance of considering multiple sources of blacklists to improve the overall performance of phishing detection.


\subsection{\textbf{ Main Objective}}

Cyber attackers can abuse social media applications' usability and functionality features to spread link-based attacks for malicious intent. We, therefore, need to better understand what security checks are in place for malicious links when shared on social apps. A key goal of this work is to pilot a standardized way of determining the adequacy of security checks against posted and shared links on popular social media applications.

\begin{itemize}
\item \textbf{RQ1 Is the user able to post a malicious link in social media applications?}\\
We want to understand if a malicious user can successfully share dangerous links. Is there a check put in place by social media applications against link-based attacks?

\item \textbf{RQ2 If the app checks for a malicious URL, can the user bypass that security check?} \\
In case it is not possible to post harmful links, is it possible for malicious users to implore the application to bypass the security check? What low-effort technique is available for the user to evade the security check put in place by social media applications?

\end{itemize}


\section{\textbf{Methodolgy}}
\label{sec:methods}

To understand the security measures in place to protect users from sharing harmful URLs on social networking apps, we employed experimental techniques in our study. We will explain the design of the study and detail how the functional components were implemented in the examination of the apps.

\subsection{\textbf{App Selection}}

\begin{table}[]
\label{tab:app_selection}
\centering
\caption{Ranking and popularity of the applications in our setup.}
\begin{tabular}{|l|l|l|}
\hline
App Name        & Installs  & Users    \\ \hline\hline
TikTok          & 1B+           & 656 M    \\ \hline
Instagram            & 1B+           & 1.21 B   \\ \hline
Twitter       & 1B+           & 429.79 M \\ \hline
Facebook             & 5B+           & 2.96 B   \\ \hline
Mostodon   & 500K+         & 4.6 M    \\ \hline 
\end{tabular}
\\ \vspace{1mm}
\footnotesize{The ranking is as of November 7, 2022, and users size is based on reports from 2021, except Mastodon which is from November 2022.  }\\
\end{table}


In this study, we selected social media applications from the Google Playstore. The sample included the top 5 social media applications in the \textit{Social} category on the Google Playstore as of November 7th, 2022. The apps included Facebook, Twitter, Instagram, TikTok, and Mastodon. These applications were selected based on their popularity and large user base, making them suitable for evaluating their security features against malicious URLs posted by users. Mastodon, despite having a relatively smaller number of users compared to the other selected applications, was included since it appeared in the top 5 \textit{Social} category of Android apps~\cite{metz_2022}. The selection of these apps was based on data of the number of installations, popularity, and user base as of the date of research.

\vspace{5pt} \noindent\textbf{Account Creation:}

In order to conduct our exploratory experiment on selected social media applications, we created fake test accounts. We used the \textit{"fakepersongenerator"} and \textit{"thispersondoesnotexist"} services to generate demographic information such as names, locations, and display pictures for these test accounts \cite{thispersondoesnotexist}\cite{fakepersongenerator}. In order to ensure the anonymity of the test accounts, we utilized the "Proton" email service that does not require mobile number verification.

\subsection{\textbf{Malicious URL Selection}}
\label{mal_url_creation}
In order to gather a set of malicious URLs, the study was conducted over a 3-month period from August 15 to November 7, 2022. This study employed a sampling methodology where 7 dates were selected with a 2-week interval between each date. For each date, 5 URLs were randomly selected from the \textit{PhishTank} database, which is a community-driven URL blocklist service that updates its database of malware and phishing links on a daily basis. The inclusion of a 2-week gap between URL selection was implemented to allow for sufficient time for social network to update their systems with the most recent malicious links reported by popular blacklist services. To further validate the malicious nature of the URLs, we utilized the services provided by \textit{Google Safe Browsing} and \textit{VirusTotal}\cite{GSB_about}\cite{VT_about}. Both services check URLs against their databases of known malicious sites and provide warnings if a URL is found to be potentially harmful. By utilizing these services, we were able to ensure that the URLs we shared on test accounts were indeed malicious.


\subsection{\textbf{URL Posting}}
\label{mal_url_posting}


We conducted the study to assess the effectiveness of malicious link detection on selected apps using test accounts. The study involved sharing malicious URLs on the apps. First, we posted the URL from \textit{Phishtank}, which we referred to as the \textit{Original} Malicious URL. If the test account was prevented from sharing the \textit{Original} Malicious URL, we created a version of the URL using \textit{Tinyurl} ,  a online service that shortens long URLs into a smaller links, which we referred to as the \textit{Redirectional} link. This \textit{Redirectional} link was then shared to test if it could bypass the detection mechanisms. We repeated this process for a total of five URLs selected from each date. This allowed us to gather a sufficient sample size to make accurate conclusions about the apps' detection capabilities.


We posted malicious URLs on test account profiles of social media platforms that allow text status updates ( such as Facebook, Twitter, and Mastodon), and sent them as direct messages on applications that did not support text sharing on user profiles with friends.


\vspace{5pt} \noindent\textbf{Ethical Considerations:}

In conducting this experiment, it was crucial to ensure that the malicious links were not accessible to actual users on the network. To achieve this, several precautionary measures were taken. Firstly, the visibility and audience of the test account were limited to the test account itself or other test accounts specifically created for this experiment. Secondly, if the \textit{Original} malicious link was not blocked by the network, no \textit{Redirectional} post was made to ensure that the link did not spread further. Lastly, all test accounts on the selected network were deleted immediately upon completion of the experimental tasks, which typically took only a few hours to complete. These steps were taken to mitigate the risk of harm to users and to ensure the integrity and validity of the experiment.

\begin{table*}[!t]
\centering
\setlength\extrarowheight{3pt}
\caption{Detail about the successfully posted URLs. }
\begin{tabular}{c|ccccc|}
\cline{2-6}
 & \multicolumn{5}{c|}{\cellcolor[HTML]{FFFFC7}\textbf{Apps}} \\ \cline{2-6} 
\textbf{Flagged Date} & \multicolumn{1}{c|}{\textbf{TikTok}} & \multicolumn{1}{c|}{\textbf{Instagram}} & \multicolumn{1}{c|}{\textbf{Twitter}} & \multicolumn{1}{c|}{\textbf{Facebook}} & \textbf{Mastodon} \\ \hline
\multicolumn{1}{|c|}{\textbf{11/7/22}} & \multicolumn{1}{c|}{5 (original)} & \multicolumn{1}{c|}{\cellcolor[HTML]{FFFFFF}5 (original)} & \multicolumn{1}{c|}{5 (only re-directional)} & \multicolumn{1}{c|}{\begin{tabular}[c]{@{}c@{}}5 (original,\\ directonal)\end{tabular}} & 5 (original) \\ \hline
\multicolumn{1}{|c|}{\textbf{10/24/22}} & \multicolumn{1}{c|}{5 (original)} & \multicolumn{1}{c|}{5 (original)} & \multicolumn{1}{c|}{3 (re-directional)} & \multicolumn{1}{c|}{4 (original)} & 5 (original) \\ \hline
\multicolumn{1}{|c|}{\textbf{10/10/22}} & \multicolumn{1}{c|}{5 (original)} & \multicolumn{1}{c|}{5 (original)} & \multicolumn{1}{c|}{1 (re-directional)} & \multicolumn{1}{c|}{5 (original)} & 5 (original) \\ \hline
\multicolumn{1}{|c|}{\textbf{9/26/22}} & \multicolumn{1}{c|}{5 (original)} & \multicolumn{1}{c|}{5 (original)} & \multicolumn{1}{c|}{2 (re-directional)} & \multicolumn{1}{c|}{\begin{tabular}[c]{@{}c@{}}5 (original,\\ re-direction)\end{tabular}} & 5 (original) \\ \hline
\multicolumn{1}{|c|}{\textbf{9/12/22}} & \multicolumn{1}{c|}{5 (original)} & \multicolumn{1}{c|}{5 (original)} & \multicolumn{1}{c|}{4 (Original, re-directional)} & \multicolumn{1}{c|}{5 (original)} & 5 (original) \\ \hline
\multicolumn{1}{|c|}{\textbf{8/29/22}} & \multicolumn{1}{c|}{5 (original)} & \multicolumn{1}{c|}{5 (original)} & \multicolumn{1}{c|}{2 (original, re-directional)} & \multicolumn{1}{c|}{5 (original)} & 5 (original) \\ \hline
\multicolumn{1}{|c|}{\textbf{8/15/22}} & \multicolumn{1}{c|}{5 (original)} & \multicolumn{1}{c|}{5 (original)} & \multicolumn{1}{c|}{4 (original, re-directional)} & \multicolumn{1}{c|}{5 (original)} & 5 (original) \\ \hline
\end{tabular}

\label{tab:posted_result}
\vspace{2mm}
\footnotesize{Each Date attribute corresponds to the date on which the randomly selected 5 URLs were flagged in PhishTank blocklist database. \textit{Redirectional} link was created using \textit{TinyURL} if the \textit{original malicious} Link selected from \textit{Phishtank} was blocked. }
\end{table*}
\begin{table}[]
\centering
\caption{Evluation Matrix for Experiment}
\label{tab:result_eval}
\begin{tabular}{|c|c|c|c|c|}
\hline
{ \textbf{App Name}} & { \textbf{Posted}} & { \textbf{Blocked}} & { \textbf{Total Attempts}} & { \textbf{Warning}} \\ \hline \hline
TikTok & 35 & 0 & 35 & 0 \\ \hline
Instagram & 35 & 0 & 35 & 1 \\ \hline
Twitter & 21 & \cellcolor[HTML]{32CB00}46 (69\%) & 67 & 0 \\ \hline
Facebook & 34 & 4(10\%) & 38 & 1 \\ \hline
Mastodon & 35 & 0 & 35 & 1 \\ \hline
\end{tabular}
\vspace{2mm}
\footnotesize{\\ A total of 23.8\% malicious links (original and transformational combined), were blocked off which Twitter had a major share. }
\end{table}




\section{\textbf{Result}}
\label{sec:result}

We examined the security measures in place on these applications to prevent users from sharing malicious links, using test accounts and flagged URLs. In the following section, we present our findings from the user's perspective on the security features of these applications. The complete overiew about each application and how it allowed and blocked original malicious links as well as their transformed version for all the selected dates can be found in the appendix table \ref{tab:complete_result_table}. In the following section, we will present an overview of the results obtained from the pilot study.

\subsection{\textbf{Malicious URL posting}}

In this section, we will describe the results from the posting of malicious URLs on social media apps in our experiment. Per Table II,  TikTok, Instagram, and Mastodon did not block any malicious links and posted all of them. Facebook blocked only  3 malicious links. We were able to post redirectional URLs for 2 of those blocked malicious links since it further blocked 1 redirection malicious link. Twitter observably blocked a major portion of the malicious URLs and their corresponding redirectional links as well. Approximately 69\% of the harmful links posted were blocked by Twitter. Table III presents the evaluation metrics of harmful links blocked by social applications. 23.8\% of the total malicious links were blocked but it was largely comprised of Twitter.

\begin{itemize}
    \item \vspace{5pt} \noindent\textbf{What is the prevalent security measure against posting malicious links?}

    Table \ref{tab:result_eval} shows results from the posting of malicious URLs for each selected application in our experiment for every chosen date. The \textit{Posted} attribute shows the number of URLs (whether original or redirectional) successfully posted, and the \textit{Blocked} row shows the number of URLs blocked. It is important to recall from \ref{mal_url_posting}, we only posted redirectional URLs if the original link was blocked.
    Despite the significant nature of the textual sharing, we observed that, with exception of Twitter, every major platform showed low or non-existent security against sharing malicious links on the platform. We can observe from Table \ref{tab:result_eval} that Twitter was the only platform that actively blocked malicious links. Facebook also blocked a few URLs. The prevalent behavior among top social network apps does not show usable security for posting features. 
    
    \item \vspace{5pt} \noindent\textbf{A few applications showed warnings when posting malicious links but did not block the links.} Instagram, Facebook, and Mastodon gave a warning for harm when we attempted to post URLs. We were still able to post these links and the applications did not remove those links, which calls into question their usable security functionality.
\end{itemize}

\subsection{\textbf{Transformation Link Posting}}
We transformed original links into redirectional links as well via a link shortening service to further our understanding of security checks for link sharing by users. We only attempted to post redirectional links for Twitter and Facebook since these were the only apps that blocked original links. On Twitter, we were not always able to pass security checks using redirectional links, particularly for the links which were a few weeks old. We observed that the security check on Facebook was relatively weaker. There were only 4 blocked dangerous links, and for each we were able to bypass the security simply using the earlier created \textit{redirectional} transformed links. Table \ref{tab:posted_result} shows complete details about successfully posted URLs.

\section{\textbf{Discussion}}
\label{sec:discuss}
\subsection{Security vs Functionality Trade-offs}
The trade-off between security and usability has been a topic of discussion in recent research on IoT devices, password managers, and mobile and web ecosystems. In system design and development, a common challenge is to balance functionality and usability with security considerations\cite{azad2019less} \cite{parkin2019usability} \cite{chaudhary2019usability}.During our study, we observed that TikTok removes the hyperlink (clickable-link) functionality from its direct messaging service when links are shared. This reduces the likelihood of malware infestation or phishing attacks by increasing the users’ effort, i.e they will have to copy and paste the link to access it.

As  discussed earlier in section \ref{sec:introduction} content sharing is an important feature in social networks, and recent literature has also discussed that E-commerce and advertising are growing on Tiktok and Dǒuyīn (TikTok’s application in mainland China) \cite{li2022critical} \cite{ma2021future}. Removing hyperlink functionality might become a design pain point for the users of the application. Security at the expense of usability, or vice-versa has been a topic of debate in recent research focusing on IoT devices, password managers, and mobile and web ecosystems. It is usually a dilemma faced in system design and development to pick functionality or usability at the cost of others. \cite{azad2019less} \cite{parkin2019usability} \cite{chaudhary2019usability}. 

\subsection{URL Blocking over Time}

For malicious link detection, it could be argued anti-phishing and malware services will take time to verify and update their databases. For example, in the case of Twitter, we observed that as we posted older links, it would block a relatively higher number of links. It is evident that in presence of a robust detection mechanism, there will be a fractional delay when the link is blocked on social media. However, as we observed from \ref{tab:result_eval} (table footnotes), that only 23.8\% of the malicious links were blocked. Given that these links were verified flagged in multiple popular anti-phishing and malware databases established that only Twitter had a substantial detection mechanism, but it was far from perfect for Facebook, and non-existent for the remaining three applications.

Due to the multipurpose functionalities offered by social media applications, the risk posed by less secure functionalities is high. Social media is primarily used for sharing and consuming content with others, and for providing updates in various forms, with textual sharing being a prominent feature. It is significant that links shared are not malicious to comply with Google Playstore safety policy and also to provide a harm-free user experience.

\subsection{Link Transformation and Security Checks}

In our experiment, we transformed links to add an intermediary point before the malicious destination link to test if the extra step in the form of link redirection will assist in bypassing the security check. According to Stivala et. al, link redirection was used to test if the platforms test the next URL in the redirection chain ~\cite{stivala2020deceptive}. Our results affirm Twitter was the only platform that proactively checked original links and actively detected redirection links as well. However, we were able to post all transformed links for blocked malicious URLs on Facebook, indicating a lack of security for redirection links in addition to original malicious links.


\section{\textbf{Future Works}}
\label{sec:future_works}
Owing to the preliminary nature of the experiment following are the limitations of this pilot study.

\begin{itemize}
\item \textbf{Investigating Security Discrepancy between Profile Posts and Private Messages}

As discussed in section \ref{sec:methods}, we would primarily post malicious URLs on profiles of test accounts.  For those platforms which do not allow textual updates on profiles, we posted the malicious links as a private direct message (DM) to another test account. The security checking behavior may differ between posts shared on the profile and direct messages, for platforms that support both functionalities. In the future, we want to investigate if there exists a security discrepancy between these two textual update functionalities against harmful link posting.

\item \textbf{Limited Visibility on Posted Malicious link}

During our experiment, we only recorded the malicious link detection behavior while posting the URLs. This was done assuming the engagement of a social media post gradually decreases over time. Recent social media marketing research sheds light on the popularity of the news posted on social networks and supports the claim of social media posts' engagement diminishing with time\cite{yang2022importance} \cite{lee2018advertising}. For future iterations, the study could be expanded to periodically check the previously posted links to see if the platforms block or delete malicious links retroactively. 

\item \textbf{Analyze Guidelines of the Google's Policy Center}

Google Malware Policy (from Play Console Help) lays out the guidelines for applications to ensure blocking dangerous links such as phishing, malware, etc. It was observed that the overall language of the policy is not strong enough for applications to block malicious URL spreading. Previous literature suggests robust policy language helps in ensuring users’ data and privacy \cite{game_priv_policy}. For future phases of the study, further understanding the security and data integration policies of social media applications.

\end{itemize}



\section{\textbf{Conclusion}}
\label{sec:conclusion}

Our study analyzed the security mechanisms of 5 social media apps from the "Socials" category on the Android Playstore against malicious link posting. This is a preliminary, systematic study that aimed to explore potential vulnerabilities related to link-based attacks on these apps. Most apps had inadequate security measures to prevent harmful links from being posted, making them vulnerable to phishing and malware attacks.  Based on these findings, all stakeholders, including social media apps and the Google Playstore, should review their security policies and implementations for secure usability.It is important to note that this study is an initial examination and that further research is needed for in-depth and detailed analysis of mitigation against link-based attacks on social network applications.

\vspace{1cm}




\bibliographystyle{plain}
\bibliography{biblio}

\section{\textbf{Appendix}}
\label{sec:appendix}
\vspace{20mm}
\begin{table*}[]
\centering
\setlength\extrarowheight{3.5pt}
\caption{Results of Original and Redirectional (\textit{Transformed}) Malicious URLs on each Application for all Selected Dates}
\label{tab:complete_result_table}
\begin{tabular}{cc|ccccc|}
\cline{3-7}
 &  & \multicolumn{5}{c|}{\cellcolor[HTML]{FFFC9E}\textbf{Apps}} \\ \cline{3-7} 
 &  & \multicolumn{1}{c|}{\cellcolor[HTML]{FFCCC9}\textbf{TikTok}} & \multicolumn{1}{c|}{\cellcolor[HTML]{FFCCC9}\textbf{Instagram}} & \multicolumn{1}{c|}{\cellcolor[HTML]{9AFF99}\textbf{Twitter}} & \multicolumn{1}{c|}{\cellcolor[HTML]{FFCCC9}\textbf{Facebook}} & \cellcolor[HTML]{FFCCC9}\textbf{Mastodon} \\ \hline
\multicolumn{1}{|c|}{} & \textbf{Posted} & \multicolumn{1}{c|}{5 (original)} & \multicolumn{1}{c|}{\cellcolor[HTML]{FFFFFF}5 (original)} & \multicolumn{1}{c|}{5 (only re-directional)} & \multicolumn{1}{c|}{\begin{tabular}[c]{@{}c@{}}5 (original,\\ directonal)\end{tabular}} & 5 (original) \\ \cline{2-7} 
\multicolumn{1}{|c|}{\multirow{-2}{*}{\textbf{11/7/22}}} & \textbf{Blocked} & \multicolumn{1}{c|}{None} & \multicolumn{1}{c|}{None} & \multicolumn{1}{c|}{5} & \multicolumn{1}{c|}{1(original)} & None \\ \hline
\multicolumn{1}{|c|}{} & \textbf{Posted} & \multicolumn{1}{c|}{5 (original)} & \multicolumn{1}{c|}{5 (original)} & \multicolumn{1}{c|}{3 (re-directional)} & \multicolumn{1}{c|}{4 (original)} & 5 (original) \\ \cline{2-7} 
\multicolumn{1}{|c|}{\multirow{-2}{*}{\textbf{10/24/22}}} & \textbf{Blocked} & \multicolumn{1}{c|}{None} & \multicolumn{1}{c|}{None} & \multicolumn{1}{c|}{\begin{tabular}[c]{@{}c@{}}7 (original +\\ re-directional)\end{tabular}} & \multicolumn{1}{c|}{2 ( 1 original and its re-direction)} & None \\ \hline
\multicolumn{1}{|c|}{} & \textbf{Posted} & \multicolumn{1}{c|}{5 (original)} & \multicolumn{1}{c|}{5 (original)} & \multicolumn{1}{c|}{1 (re-directional)} & \multicolumn{1}{c|}{5 (original)} & 5 (original) \\ \cline{2-7} 
\multicolumn{1}{|c|}{\multirow{-2}{*}{\textbf{10/10/22}}} & \textbf{Blocked} & \multicolumn{1}{c|}{None} & \multicolumn{1}{c|}{None} & \multicolumn{1}{c|}{\begin{tabular}[c]{@{}c@{}}9 (original +\\ re-directional)\end{tabular}} & \multicolumn{1}{c|}{None} & \begin{tabular}[c]{@{}c@{}}Warning for one \\ but not blocked\end{tabular} \\ \hline
\multicolumn{1}{|c|}{} & \textbf{Posted} & \multicolumn{1}{c|}{5 (original)} & \multicolumn{1}{c|}{5 (original)} & \multicolumn{1}{c|}{2 (re-directional)} & \multicolumn{1}{c|}{5 (original + re-direction)} & 5 (original) \\ \cline{2-7} 
\multicolumn{1}{|c|}{\multirow{-2}{*}{\textbf{9/26/22}}} & \textbf{Blocked} & \multicolumn{1}{c|}{None} & \multicolumn{1}{c|}{None} & \multicolumn{1}{c|}{\begin{tabular}[c]{@{}c@{}}8 (original +\\ re-directional)\end{tabular}} & \multicolumn{1}{c|}{1 (original)} & None \\ \hline
\multicolumn{1}{|c|}{} & \textbf{Posted} & \multicolumn{1}{c|}{5 (original)} & \multicolumn{1}{c|}{5 (original)} & \multicolumn{1}{c|}{\begin{tabular}[c]{@{}c@{}}4 ( original + \\  re-directional)\end{tabular}} & \multicolumn{1}{c|}{5 (original)} & 5 (original) \\ \cline{2-7} 
\multicolumn{1}{|c|}{\multirow{-2}{*}{\textbf{9/12/22}}} & \textbf{Blocked} & \multicolumn{1}{c|}{None} & \multicolumn{1}{c|}{\begin{tabular}[c]{@{}c@{}}1 phishing warning\\ (not blocked)\end{tabular}} & \multicolumn{1}{c|}{6 blocked} & \multicolumn{1}{c|}{\begin{tabular}[c]{@{}c@{}}1 phishing warning but\\ not blocked\end{tabular}} & None \\ \hline
\multicolumn{1}{|c|}{} & \textbf{Posted} & \multicolumn{1}{c|}{5 (original)} & \multicolumn{1}{c|}{5 (original)} & \multicolumn{1}{c|}{\begin{tabular}[c]{@{}c@{}}2 (original + \\  re-directional)\end{tabular}} & \multicolumn{1}{c|}{5 (original)} & 5 (original) \\ \cline{2-7} 
\multicolumn{1}{|c|}{\multirow{-2}{*}{\textbf{8/29/22}}} & \textbf{Blocked} & \multicolumn{1}{c|}{None} & \multicolumn{1}{c|}{None} & \multicolumn{1}{c|}{7 blocked} & \multicolumn{1}{c|}{None} & None \\ \hline
\multicolumn{1}{|c|}{} & \textbf{Posted} & \multicolumn{1}{c|}{5 (original)} & \multicolumn{1}{c|}{5 (original)} & \multicolumn{1}{c|}{\begin{tabular}[c]{@{}c@{}}4 (original +\\ re-directional)\end{tabular}} & \multicolumn{1}{c|}{5 (original)} & 5 (original) \\ \cline{2-7} 
\multicolumn{1}{|c|}{\multirow{-2}{*}{\textbf{8/15/22}}} & \textbf{Blocked} & \multicolumn{1}{c|}{None} & \multicolumn{1}{c|}{None} & \multicolumn{1}{c|}{4 blocked} & \multicolumn{1}{c|}{None} & None \\ \hline
\end{tabular}
\vspace{3mm}
\footnotesize{ \\ Each Date attribute corresponds to the date on which the randomly selected 5 URLs were flagged in the PhishTank blocklist database. \\ We observed that only Twitter had a robust security check put in place to proactively block malicious links.}
\end{table*}

\end{document}